 \documentclass[final,5p,times,twocolumn]{elsarticle}
\usepackage[colorlinks=true, allcolors=blue]{hyperref}

\usepackage{amssymb}
\usepackage{amsmath}
\usepackage{siunitx}
\DeclareSIUnit\barn{b}
\usepackage{tabularx}

\usepackage{dcolumn}% Align table columns on decimal point
\usepackage{placeins}
\usepackage{subcaption}
\usepackage{orcidlink}

\usepackage{tikz}
\tikzset{
  level/.style   = { ultra thick, blue, font=\small },
  connect/.style = { dashed, font=\small },
  notice/.style  = { draw, rectangle callout, callout relative pointer={#1}, font=\small },
  label/.style   = { text width=2cm, font=\small }
}
\usetikzlibrary{patterns}
\tikzstyle{EDR}=[draw=none,line width=1pt,preaction={clip, postaction={pattern=north west lines, pattern color=black}}]
\usepackage[dvips]{xcolor}
\usepackage{nicefrac}

\newcommand{\mycomment}[1]{}

\journal{European Physical Journal D}

\begin{document}

\begin{frontmatter}



\title{Collinear Laser Spectroscopy on a Fast Atomic Beam of Boron \\ Generated by Photodetachment of Accelerated B$^{-}$ Ions}

\author[IKP]{L. Renth\,\orcidlink{0000-0003-0879-1751}}
\author[IKP]{P. Imgram\,\orcidlink{0000-0002-3559-7092}}
\author[IKP]{B. Maass\,\orcidlink{0000-0002-6844-5706}}
\author[UG]{D. Hanstorp\,\orcidlink{0000-0001-6490-6897}}
\author[IKP,HFHF]{J. Krämer\,\orcidlink{0000-0001-7775-7376}}
\author[IAP]{D. Koestel\,\orcidlink{0000-0001-6059-6254}}
\author[UG]{D. Lu\,\orcidlink{0000-0002-2439-7124}}
\author[IKP,HFHF]{W. Nörtershäuser\,\orcidlink{0000-0001-7432-3687}}
\author[IAP,HFHF]{T. Walther\,\orcidlink{0000-0001-8114-1785}}
\affiliation[IKP]{organization={Institut für Kernphysik, Technische Universität Darmstadt},
%    addressline={Schloßgartenstraße 9},
    city={Darmstadt},
    postcode={64289},
%%           state={},
    country={Germany}}
\affiliation[UG]{organization={Department of Physics, University of Gothenborg},
    country={Sweden}}

\affiliation[IAP]{organization={Institut für Angewandte Physik, Technische Universität Darmstadt},
%    addressline={Hochschulstr. 6},
    city={Darmstadt},
    postcode={64289},
%%           state={},
    country={Germany}}

\affiliation[HFHF]{organization={Helmholtz Forschungsakademie Hessen für FAIR (HFHF), GSI Helmholtzzentrum für Schwerionenforschung GmbH},
%    addressline={Planckstr. 1},
    city={Darmstadt},
    postcode={64291},
%%           state={},
    country={Germany}}

\begin{abstract}
%\input{abstract}
Atomic beams for collinear laser spectroscopy are typically produced via charge exchange reactions in an in-beam vapor cell. This process is accompanied by the formation of a considerable amount of longer-lived excited state population, which is not accessible for spectroscopy and also induces fluorescence background to photon detectors. We present an alternative method to produce an atomic beam, consisting exclusively of ground-state atoms. Negative ions are neutralized in-flight by photodetachment and are subsequently used for fluorescence spectroscopy. As a test candidate, a negative boron ion beam was produced in a cesium sputtering source and superimposed with a co-propagating high-power pulsed infrared laser. The neutral atoms were subsequently excited along the $2s^2\,2p\,^{2}\mathrm{P}_{\nicefrac{1}{2},\nicefrac{3}{2}} \rightarrow 2s^2\,3s\,^{2}\mathrm{S}_{\nicefrac{1}{2}}$ transitions by a continuous-wave laser at about 250\,nm. The statistics and efficiency were limited by the combination of a continuous ion beam with a low-repetition-rate laser, but a clear route for the improvement combining existing techniques is presented, which can then be applied to other elements as well as negative molecular ions.
\end{abstract}




\end{frontmatter}

%\tableofcontents



\section{Introduction}
\label{introduction}
Collinear laser spectroscopy (CLS)~\cite{Kaufman.1976,Wing.1976,Meier1977} is a renowned tool for investigating the electronic structure of ions~\cite{Meier1977,Silverans.1980,Mueller.1983}, atoms~\cite{Anton.1978,Schinzler.1978}, and molecules~\cite{Wing.1976,Carrington.1979,Hechtfischer.1998,GarciaRuiz.2020} and for extracting nuclear ground-state properties from optical spectra~\cite{Cheal.2010,KlausBlaum.2013,Campbell.2016,Neugart.2017,Yang.2023}. For CLS, the ions of interest are accelerated to form an ion beam, which is then superimposed with a narrow-bandwidth laser beam. The electrostatic acceleration of the ions results in a compression of the velocity distribution, which consequently leads to a reduced Doppler width of the optical transitions \cite{Kaufman.1976}. An ion energy of 20\,keV reduces the Doppler width by approximately two orders of magnitude, thus enabling the observation of resolved hyperfine spectra of atomic and ionic transitions. Rather than scanning the laser frequency, it is customary to apply an additional post-acceleration voltage to match the Doppler-shifted frequency to the atom's resonance frequency. The subsequent emission of photons can be measured using photomultiplier tubes (PMTs) but a multitude of other detection methods can be combined with CLS to obtain higher efficiency~\cite{Cheal.2010,KlausBlaum.2013,Campbell.2016,Neugart.2017,Yang.2023}.  

The most direct case is the use of transitions from the ionic ground state, given that the majority of ion sources deliver exclusively or at least dominantly such ions. Consequently, a large fraction of the ions in the beam can be addressed and will contribute to the signal. However, ionic ground-states often do not have laser-accessible ($\lambda>\SI{200}{nm}$) and fast ($\tau < \SI{e-7}{s}$) optical transitions, while the ground or metastable state in the corresponding neutral atomic system does. In such cases, the conventional approach utilizes a charge-exchange cell, which is filled with alkaline vapor, to effectively neutralize the incoming ion beam \cite{Anton.1978,Klose.2012}. But the population distribution across atomic states becomes extensive in the charge-exchange process  \cite{Vernon.2019}, thereby diminishing the efficiency. Moreover, electrons ending up in excited states will cascade down to the ground state, inducing background in the fluorescence detection system. Finally, inelastic collisions between beam and vapor atoms induce a velocity shift that can give rise to an asymmetry or additional side peaks in the optical spectrum  \cite{Bendali.1986, Neugart.1977,Klose.2012}. Thus, the charge-exchange approach can deteriorate both the sensitivity and the accuracy of collinear spectroscopy.

Previously, we have demonstrated CLS measurements with accuracy at the \SI{100}{\kHz} level for Ba$^+$~\cite{Imgram.2019} and Ca$^+$~\cite{PatrickMuller.2020} ions using the Collinear Apparatus for Laser Spectroscopy and Applied Physics (COALA)~\cite{KKoening.2020}. COALA was not equipped with a charge-exchange cell, and the collisional charge-exchange process is in general plagued by line-shape issues. Consequently, measurements utilizing charge-exchange cells with systematic uncertainties $\leq \SI{1}{\MHz}$ and statistical uncertainties $\leq \SI{1}{\MHz}$ have so far only been reported for the $^{19,21}$Ne isotope shift~\cite{Marinova2011}.

An illustrative case where such accuracy is still lacking is boron. The boron isotopes are of particular interest for laser spectroscopy since $^8$B is believed to be the prototype of a proton-halo nucleus, and the determination of its charge radius would provide an important benchmark for ab initio nuclear structure theory (see, \textit{e.g.},~\cite{Carlson.2015}). Such a measurement is currently being prepared at the ATLAS facility at Argonne National Laboratory~\cite{Maa.2017}. Even for the stable isotopes, neither the charge radius nor the differential charge radii were well known before the precise measurement of the $2s^2\,2p\; ^2\mathrm{P}_{\nicefrac{1}{2}} \rightarrow 2s^2\,3s\; ^2\mathrm{S}_{\nicefrac{1}{2}}$ transition in $^{10,11}$B was performed using resonance ionization mass spectroscopy on a thermal atomic beam~\cite{Maa.2019}. The results were analyzed using novel mass-shift calculations in the five-electron system and compared to ab initio nuclear-model calculations. Since the calculated mass shift -- a required input parameter for the charge-radius determination -- is still more than an order of magnitude more precise than the experimental result, the uncertainty in the charge-radius difference would directly benefit from a more accurate measurement. This makes boron a compelling but demanding test case, calling for a clean, well-controlled beam-preparation method as an alternative to charge exchange.

Here, we present an alternative approach for CLS that produces an atomic beam in its ground or metastable state without these drawbacks. We start with a negative ion beam and neutralize the ions in flight by photodetachment in front of and within the optical detection region. Laser photodetachment has been applied to various elements, such as He \cite{Kristensen.1997}, carbon \cite{Bresteau2016}, iodine \cite{Rothe.2017}, and astatine \cite{Leimbach.2020}, to determine the electron affinity of the respective species. It has recently been combined with subsequent resonance ionization as an efficient detection tool~\cite{Welander.2022} and was also applied in a multi-reflection time-of-flight (MR-ToF) spectrometer~\cite{maier_enhanced_2025}.

Beyond its use in precision spectroscopy, photodetachment has also become an established technique for producing fast beams of neutral atoms for a wide range of applications. In fusion research, photodetachment of H$^-$ or D$^-$ is being pursued as a clean alternative to gas-cell neutralization for the heating neutral-beam injectors planned for future reactors such as ITER and DEMO \cite{Blondel2019, Simonin2015, Simonin2021}. Continuous-wave photoneutralization of a H$^-$ beam has been demonstrated to saturate at high efficiency using an enhancement cavity, a key step towards practical photon-based neutral-beam injectors~\cite{Bresteau2017}. 
At the Cryogenic Storage Ring (CSR) in Heidelberg, fast neutral-atom beams generated by laser photodetachment of negative ions are merged with stored molecular ions to study ion--atom collision and reaction dynamics at low relative energies~\cite{OConnor2015, Becker2022}.
Photodetachment of negative hydrogen-like ions also plays a central role in antimatter research: in the GBAR experiment at CERN, cold antihydrogen atoms are produced by photodetaching the outer positron from trapped, laser-cooled $\overline{\mathrm{H}}^+$ ions, enabling a free-fall measurement of the gravitational acceleration of
antimatter~\cite{Debu2013, Perez2015}. In a related application, a photodetached neutral hydrogen beam was merged with a Si$^{3+}$ ion beam to measure absolute cross sections for low-energy charge transfer between multiply charged ions and atomic hydrogen~\cite{Bruhns2008}. These diverse applications illustrate that photodetachment provides a versatile and clean route to well-defined neutral atomic beams, motivating its use here for high-resolution collinear laser spectroscopy.

In this work, we apply this process specifically to prepare an atomic beam \emph{in its ground state} -- as opposed to the excited or mixed-state beams typically produced by collisional charge exchange -- which is essential for the high-precision laser spectroscopy required for cases like boron discussed above. In our proof-of-principle experiment, we applied it to boron ions and investigated the  $2s^2\,2p\; ^2\mathrm{P}_{\nicefrac{1}{2},\nicefrac{3}{2}} \rightarrow 2s^2\,3s\; ^2\mathrm{S}_{\nicefrac{1}{2}}$ transitions in neutral $^{10,11}$B.

\section{Methods}\label{sec:methods}

\begin{figure*}[!h]
    \centering
    \includegraphics{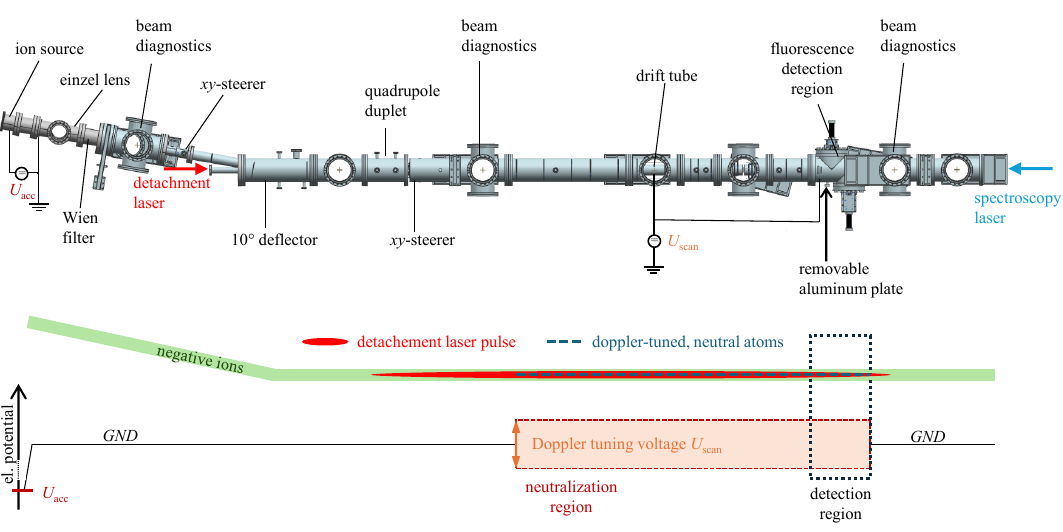}
    \caption{Top: The Collinear Apparatus for Laser Spectrcoscopy and Applied Physics (COALA). Ions are accelerated with $U_\mathrm{acc}$ into the beamline from the left and can be detected on a first beam diagnostic station. The ion beam is focused by an einzel lens and guided with $xy$-steerers and a 10°-deflector into the spectroscopy beamline where it can be superimposed with copropagating and counterpropagating laser beams using $xy$-steerers and a quadrupole duplet. The beams can be monitored on the second and third beam diagnostic stations to ensure their superposition. The fluorescence detection region and the preceding drift tube can be put on an additional voltage $U_\mathrm{scan}$ for Doppler tuning. In this work, a removable aluminum plate was added to the detection region, allowing the detection of arriving particles by impact fluorescence. Bottom: Acceleration and neutralization scheme. Negative ions are accelerated from a negative potential $U_\mathrm{acc}$ to ground potential (\textit{GND}). The detachment laser is superimposed with the ion beam. The negative ions are (de-)accelerated by the voltage $U_\mathrm{scan}$ applied to the drift tube and detection region. Only ions within this neutralization region are Doppler tuned for laser spectroscopy before they are neutralized by the detachment laser pulses. }
    \label{fig:COALA}
\end{figure*}

The proof-of-principle measurements were performed at the COALA beamline \cite{KKoening.2020} in the Institute for Nuclear Physics of the Technical University Darmstadt. The upper part of Fig.\,\ref{fig:COALA} shows the principle setup of COALA. Ions are produced on the left side and accelerated by a high voltage of typically \SI{20}{\kilo\volt} into the beamline. Several steerers, a 10°-deflector, an einzel lens and a quadrupole doublet are used to shape the ion beam and to steer it straight through the beamline and the fluorescence detection region. To monitor the beam, three beam diagnostic stations are equipped with Faraday cups to read the ion current, phosphor screens to investigate the shape of the beam, and irises which can be closed and opened to localize the ion and the laser beam and facilitate their superposition. The fluorescence detection region consists of two elliptical mirror systems that focus the light emitted from the beam axis to a set of photomultiplier tubes (PMTs) \cite{maas_4pi_2020}. An additional voltage can be applied to the detection region and the preceding drift tube for Doppler tuning.
Laser beams can be coupled into the beamline so that they propagate either with or against the ion beam. More details on COALA can be found in \cite{KKoening.2020}. In the following, we address specific points that had to be implemented to perform the proof-of-principle experiment.

\subsection{Negtive Ion Production}\label{sec:source}
A cesium sputtering ion source \cite{Middleton.1983} was installed to produce a beam of B$^-$ ions. 
Cesium is vaporized from a reservoir and fed into the source body. An ionizer anode on a positive potential with respect to the cathode surface ionizes the cesium vapor. The cathode contains the target material of interest, often mixed with some other material to improve the sputtering process. In this case it is made of a copper cylinder filled with a mixture of pressed silver and boron powder as suggested by Middelton \cite{Middleton.1989}. The potential difference between ionizer and cathode accelerates the positive cesium ions onto the target.

The cathode and target are cooled such that cesium can condensate at the target surface. The  accelerated cesium ions sputter out material from the target and in combination with the condensed cesium layer, acting as a charge exchange medium, B$^-$ ions are produced and accelerated away from the sputter target and extracted towards ground potential into the collinear beamline. The source \cite{Blahins.2020} was operated at a total acceleration voltage of \SI{6}{\kV} and it was placed on a floatable potential of up to \SI{20}{\kV} as higher  higher kinetic energy gives lower emittance. Further, a higher beam energy is advantageous for CLS.  

\subsection{Time-of-Flight Measurements}

\begin{figure}[!h]
    \centering    \includegraphics{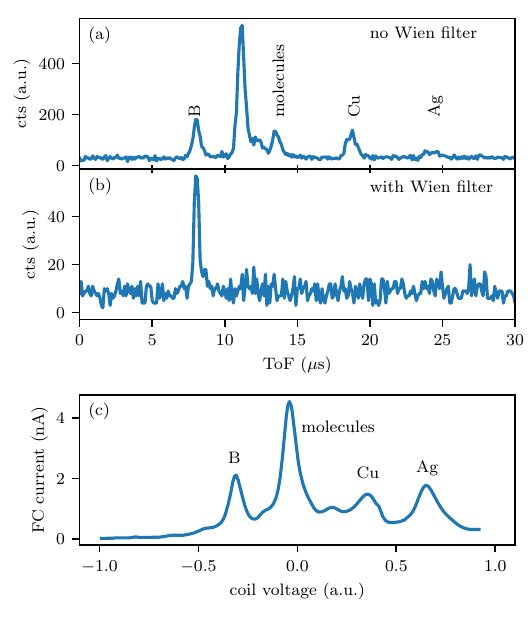}
    \caption{Time-of-flight spectra of the negative ion beam from the sputter source (a) without and (b) with mass separation by a Wien filter installed between the ion source and the $10^\circ$-deflector. Both spectra were recorded at $\SI{24}{\kilo e \volt}$ beam energy and with the continuous ion beam chopped to $\SI{300}{\nano\second}$ long pulses at the deflector. In (a), the peaks with increasing ToF are assigned as $^{10,11}$B, molecular clusters, copper and silver ions. In (b) the effect of purification with the Wien filter is shown: all higher masses than boron are removed from the beam. The measurement time of this spectrum was only $\approx \nicefrac{1}{3}$ of the measurement time of the upper spectrum, \textit{i.e.}, the transmission of boron ions through the Wien filter is almost 100\%. (c) Wien-filter spectrum at a constant Wien-filter electric field of 300\,V. The $x$-axis represents the voltage applied to the coils that induce the magnetic field of the Wien filter. The identified peaks are labeled.
    }
    \label{fig:ToF}
\end{figure}

The negative ion sputtering source with a copper-silver-boron target produces many different kinds of charged ions and molecules \cite{Middleton.1989}. Therefore, the actual beam composition was investigated in a first step. The COALA beamline is not equipped with a mass-separating magnet to identify the different masses in the ionic beam. Instead, we have implemented a time-of-flight (ToF) technique that allows us to investigate ion beams delivered by a continuous ion source. The potential on the first horizontal steerer was pulsed by a fast high-voltage solid-state switch in order to chop the continuous ion beam. The pulse length was set to $\SI{300}{\nano\second}$, \textit{i.e.}, the ions were deflected into the beamline only for $\SI{300}{\nano\second}$ and efficiently blocked for the remaining time. 
The fluorescence detection region ($\SI{5\pm.1}{\meter}$ after the $xy$-steerer) was modified and a small aluminum plate was moved into the beam. Impinging ions cause a light flash which is detected by the PMTs, constituting a very simple and fast particle detector. 

\begin{figure}[!h]
    \centering
    \includegraphics{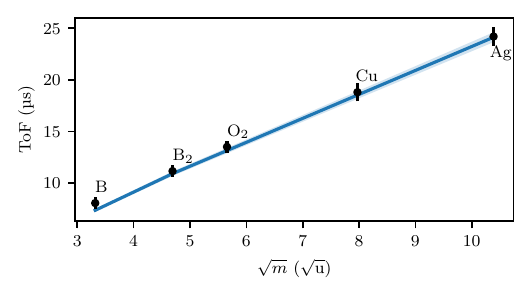}
    \caption{The observed ToFs of the different components in the negative ion beam are plotted against the square root of their atomic masses. Error bars represent the FWHM of the observed ToF peaks. The solid line is the expected ToF-mass relation. 
    %The measurements and calculations were performed for $\SI{24}{\kilo e \volt}$ kinetic energy and a traveling distance of $\SI{5.0\pm.1}{\meter}$. 
    The uncertainties of the expected values are propagated from the uncertainty in the traveling distance and represented as a shaded area.
    \label{fig:ToF_m}}
\end{figure}

Triggering the timer for the signal detection with the pulse for the high-voltage switch, the  ToF of the ions between the deflector and the detection region is measured, allowing an extraction of the ions' $m/q$. Figure\,\ref{fig:ToF} shows two ToF spectra recorded at $\SI{24}{\kilo e \volt}$ beam energy. The upper spectrum (a) shows the ToF spectrum of the full beam from the sputter source. From the observed peaks, the flight time of the different beam components were extracted and identified. Their values are plotted in Fig.\,\ref{fig:ToF_m} together with the ToF-mass relation expected at $\SI{24}{\kilo e \volt}$ and a traveling distance of $\SI{5\pm0.1}{m}$ from the steerer to the detection region. 
The observed ToFs are all slightly larger than the theoretical predictions, which might indicate either a slightly smaller beam energy, but are still in very good agreement with the expected values.

The various elements and molecules in the beam impede a proper ion tuning on $^{10}$B or $^{11}$B and lead to an increased background rate in the detection region. Therefore, we installed a Wien filter between the ion source and the 10°-deflector to purify the beam. A mass spectrum taken with the Wien filter is shown in Fig.\,\ref{fig:ToF}(c) and the same ion species are identified. Adjusting the magnet coil voltage to the maximum of the boron peak, the other beam components are efficiently removed from the beam without a significant loss of B$^-$ ions as documented in the ToF spectrum in Fig.\,\ref{fig:ToF}(b). We note that the spectrum in (b) was accumulated for only one third of the time of the spectrum in (a). The Wien filter now allows an appropriate beam tuning to ensure that the laser beam is superimposed with the right ion species.

\subsection{Photodetachment}\label{sec:pd}
\begin{figure}[h]
    \centering
    \begin{tikzpicture}[
            scale=0.45,
            font=\footnotesize,
            level/.style={thick},
            virtual/.style={thick,densely dashed},
            trans/.style={thick,<->,shorten >=2pt,shorten =2pt,>=stealth},
            classical/.style={thin,double,<->,shorten >=4pt,shorten <=4pt,>=stealth}
        ]
       \draw[level] (0,0) -- node[above] {B$^-$ $2s^2\,2p^2\, ^3$P} (5,0);
       \node[below] at (0,0) {-0.28 eV};
       \draw[connect] (5,0) -- (6,0.6) (5,0) -- (6,-0.5) (5,0) -- (6, -1.2);
       \draw[level] (6,0.6) -- (7.5,0.6) node[right]{$J$=2};
       \draw[level] (6,-0.5) -- (7.5,-0.5) node[right] {$J$=1};
       \draw[level] (6,-1.2) -- (7.5,-1.2) node[right] {$J$=0};
       \draw[level] (10,3)  -- (12.5, 3) node[below] {$J$=1/2};
       \draw[level] (10,4)  -- (12.5, 4)node[above] {$J$=3/2};
       \draw[connect] (13.5,3.5) -- (12.5,4) (13.5,3.5) -- (12.5,3);
       \draw[level] (13.5,3.5) -- node[above] {B $2s^2\,2p \,^2\mathrm{P}$} (17.5, 3.5);
       \draw[EDR] (6, 3) rectangle (9.5, 4);
       \draw[level] (6, 3)  -- (9.5, 3);
       \draw[fill=white, draw=none] (5.9, 3.5) rectangle (9.6, 4.1);
       \node[below] at (9, 3) {0 eV};
       \draw[->] (6.5,0.6) -- node [above] {} (6.6, 6.4);
       \draw[->] (6.7,-0.5) -- node [above] {} (6.7, 6.2);
       \draw[->] (6.9,-1.2) -- node [above] {} (6.8, 6);
       \draw[dashed, ->] (6.8, 6.4) -- (11, 3.1);
       \draw[dashed, ->] (6.8, 6.4) -- (10.2, 4.1);
       \draw[level] (10, 7) -- node[above] {J=1/2 \ \ \ \ B $2s^2\, 3s\,^2\mathrm{S}$} (17.5, 7);
       \node[below] at (16.5, 7) {$\SI{4.96}{e\volt}$};
       \draw[->] (11, 3) -- (11, 6.9);
       \draw[->] (10.2, 4) -- (10.2, 6.9);
    \end{tikzpicture}
    \caption{Energy level scheme (not to scale) of B$^-$ (left) and neutral B (right). The solid arrows indicate the possible optical transitions by photodetachement in negative boron and the resonant optical transitions in neutral boron. The dashed arrows indicate decays through electron emission.}
    \label{fig:levels}
\end{figure}
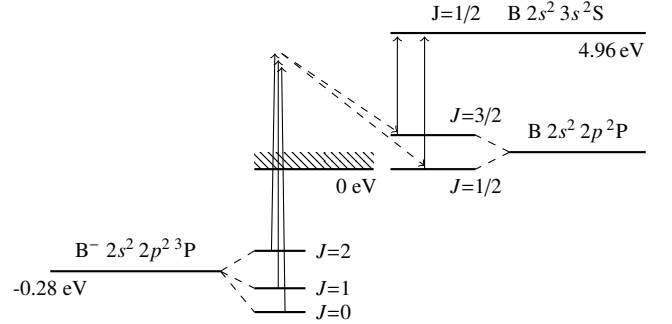

In the photodetachment process of a negative ion, a photon is absorbed by the ion, followed by the emission of an electron, leaving the neutral atom in an energetically allowed excited state or in its ground state. 
For photodetachment in B$^-$, a transition energy of \SI{0.28}{\eV} \cite{Scheer.1998} is required, which corresponds to a wavelength of $\SI{4.4}{\micro\meter}$ or lower. In our proof-of-principle experiment, we used a pulsed infrared laser with a wavelength of \SI{1086.6}{\nano\meter} ($E_\gamma=\SI{1.14}{\eV}$) that was developed for a Brillouin-LIDAR experiment \cite{Rupp.2018}. The photon energy is, thus, sufficient to populate both $^2\mathrm{P}_{\nicefrac{1}{2},\nicefrac{3}{2}}$ fine structure components of the atomic ground state, which are only separated by about \SI{1.9}{\milli e\volt}. The energy level schemes of B$^-$ and neutral boron are depicted in Fig.\,\ref{fig:levels} and the possible transitions are indicated. At $\lambda = \SI{1086.6}{\nano\meter}$, a total photodetachment cross section of $\sigma\approx\SI{3E-17}{\centi\meter\squared}$ is estimated \cite{Liu.2013}.  The detachment laser provides pulse energies of up to $E_\mathrm{pulse}=\SI{350}{\micro J}$ and exhibits a pulse length of \SI{12}{\nano \second} at a repetition rate of \SI{1}{\kilo \hertz}. Together with its beam diameter of $d=2r=\SI{5}{\milli\meter}$, the total neutralization efficiency with a perfect ion and laser beam overlap can be estimated to
\begin{equation}
    \frac{n_\mathrm{B}}{n_\mathrm{B^-}}=\sigma \; \frac{n_\mathrm{B^-}}{\pi r^2}\frac{n_{\gamma}}{n_\mathrm{B^-}}=\sigma \; \frac{E_\mathrm{pulse}}{h\nu \; \pi r^2} = \SI{29}{\%}
    \label{equ:eff}
\end{equation}
where $n_\mathrm{B}$ is the number of neutralized atoms, $n_\mathrm{B^-}$ the number of ions traveling through a laser pulse and $n_{\gamma}$  the number of photons in this pulse. 
%\textcolor{red}{What was the real beam diameter of the pulsed laser?}

The detachment laser was overlapped with the copropagating ion beam in order to achieve a large interaction volume and best neutralization rates. The relative strengths of the neutralization transitions given in \cite{Scheer.1998} predict that about $\nicefrac{2}{3}$ of photodetachment processes lead to the $^2$P$_{\nicefrac{3}{2}}$ state and $\nicefrac{1}{3}$ to the $^2$P$_{\nicefrac{1}{2}}$ state, which is in excellent agreement with the expected distribution from the statistical argument of the available fine-structure states. 
%\textit{i.e.}, partial neutralization efficiencies of $\SI{19.3}{\percent}$ into the $^2$P$_{\nicefrac{3}{2}}$ state and $\SI{9.3}{\percent}$ into the $^2$P$_{\nicefrac{1}{2}}$ state for perfect laser and ion beam overlap.

The  detachment laser interacts with the ions along $\approx\SI{4.5}{\meter}$ between the $\SI{10}{\degree}$-deflector and the end of the detection region. At $\SI{17}{\kilo e \volt}$ (which resulted in the most stable beam properties), the ions need about $\SI{8}{\micro\second}$ to travel the complete distance, \textit{i.e.}, the whole neutralization area is refilled with negatively charged ions from the continuous ion beam every $\SI{8}{\micro\second}$ after an IR laser pulse. The highest neutralization efficiency could therefore be reached with a repetition rate of the pulsed laser of $\SI{125}{\kilo\hertz}$. During the experiment, a repetition rate of only $\SI{1}{\kilo\hertz}$ was available, reducing the temporal efficiency by more than two orders of magnitude.

\subsection{Collinear Laser Spectroscopy on Photodetached Boron}\label{sec:colaspec}

For the proof-of-principle experiment, the laser beam for photodetachment was collinearly superimposed with the continuous ion beam and operated at about \SI{1}{\kHz} with a typical pulse energy of about $\SI{300}{\micro\joule}$, which corresponds to an average laser power of $\SI{300}{\milli\watt}$. The beam diameter of the pulsed laser inside the beamline was $\SI{5}{\milli\meter}$. 

For the resonant excitation of the produced boron atoms, a continuous wave (cw) laser with frequency close to the Doppler-shifted transition frequency of the neutral atom was adjusted counterpropagating to the ion beam and the photo-detachment laser. The cw laser has a beam diameter of $d_s = \SI{2}{\milli\meter}$ and its power was only about $\SI{1}{\milli\watt}$ before entering the beam line. The difference of laser beam diameters reduces the efficiency by a factor of 0.16. 

The drift tube and the detection region were set on an additional voltage (see Fig.\,\ref{fig:COALA}), which was scanned between $-\SI{50}{\volt}$ and $+\SI{50}{\volt}$ for Doppler-tuning of the ion velocity before neutralization. With this configuration, ions along a total length of about $\SI{3.2}{\meter}$ were neutralized at the correct velocity for laser excitation. This increases the interaction length of the detachment laser with the Doppler-tuned ion beam, but has the disadvantage that atoms upstream in the  drift tube can already be excited by the spectroscopy laser along the $\SI{3.2}{\meter}$ length. The resulting possibility for population transfer into another fine-structure or hyperfine-structure level before reaching the detection region effectively decreases the detectable fluorescence light inside the detection region.

\begin{figure}[h]
    \centering
    \includegraphics{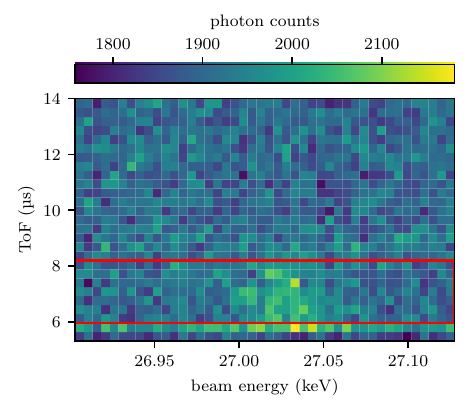}
    \caption{Time resolved spectrum of $^{11}$B. The horizontal axis represents the ion beam energy in $\si{keV}$ and the vertical axis the time in $\si{\micro\second}$ between the trigger signal and the detection signal. The detected photon counts are color-coded. Ions of different initial velocities require different Doppler-tuning voltages to show resonant fluorescence. Photons from the detachment laser pulse are detected shortly before $\SI{6}{\micro\second}$. The resonance is visible for about $\SI{2}{\micro\second}$ after the detachment laser pulse. Only data within this time gate (marked with red lines) is used for the analysis.}
    \label{fig:trs}
\end{figure}

For the time-resolved data acquisition, the arrival times of the PMT signals were measured with respect to the trigger of the detachment laser.
Figure \ref{fig:trs} shows a time-resolved spectrum taken on $^{11}$B within about \SI{40}{\minute}. 
We note that this corresponds to $\approx\SI{2.4E6}{}$ detachment-laser pulses or \SI{4.8}{\second} 
of neutralized beam, resulting in a temporal efficiency of \SI{2E-3}{}. The tail of the detachment laser pulse is still visible at the lower part of the spectrum, before $\SI{6}{\micro\second}$, as an increased photon count along the complete voltage axis. The resonance then stays visible for $\SI{2}{\micro\second}$ after the detachment-laser pulse, which -- taking the ion velocity into account -- leads to an estimate of an effective neutralization-length of roughly $\SI{1.1}{\meter}$. By time-gating on these $\SI{2}{\micro\second}$, as indicated in the figure by the horizontal lines, the high background signal from the detachment laser as well as 99.8\% of the cw laser's background are removed from the spectrum.

\begin{table}
    \centering
\caption{Summary of efficiencies for neutral beam production in the proof-of-principle experiment. Not included are the population distribution to different fine-structure and hyperfine-structure states. For details see text.}
\label{tab:efficiencies}
    \begin{tabular}{ll}
    \hline\hline
         detachment efficiency       & 0.29\\
         temporal overlap            & \SI{2E-3}{}\\
         spatial overlap laser beams & 0.16\\
    \hline
    \rule{0mm}{3.5mm}     total& \SI{9E-5}{}\\
    \hline\hline
    \end{tabular}
\end{table}

\section{Results and Discussion}
With the new developments at COALA discussed in Sec.\,\ref{sec:methods}, collinear laser spectroscopy was performed on neutral boron produced by photodetachment from B$^-$ ions.

During the measurement, an ion current of \SI{2.47}{\nano\ampere} was monitored at the end of the beamline with both irises set on $\SI{5}{\milli\meter}$ and with the Wien filter on. 
%The ToF of boron through the complete drift tube and detection region is about $\SI{5}{\micro\second}$. 
Taking into account the efficiencies estimated in the previous section and summarized in Tab.\,\ref{tab:efficiencies}, as well as the natural abundances of 80\% for $^{11}$B and  20\% for $^{10}$B, this corresponds to about 1200 atoms of $^{11}$B per laser pulse, of which 800 are estimated to be in the $^2\mathrm{P}_{\nicefrac{3}{2}}$ and 400 in the $^2\mathrm{P}_{\nicefrac{1}{2}}$ level. These atoms are additionally distributed across the different hyperfine states. The total flux of neutral atoms corresponds to about \SI{200}{\femto\ampere} for $^{11}$B and \SI{50}{\femto\ampere} of $^{10}$B .  

\begin{figure*}[ht]
    \begin{subfigure}[t]{0.33\textwidth}
    %\caption{$^{11}$B, from $^2\mathrm{P}_{\nicefrac{1}{2}}$}
    \centering
        \includegraphics{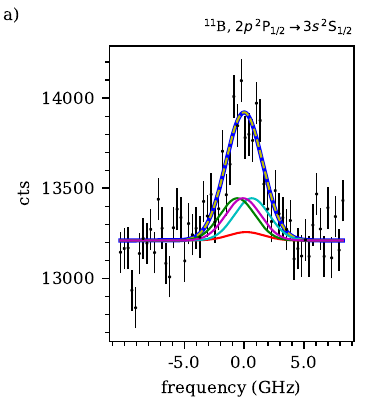}
    \label{fig:11BD1res}
    \end{subfigure}
    \begin{subfigure}[t]{0.33\textwidth}
    %\caption{$^{11}$B, from $^2\mathrm{P}_{\nicefrac{3}{2}}$}
    \centering
        \includegraphics{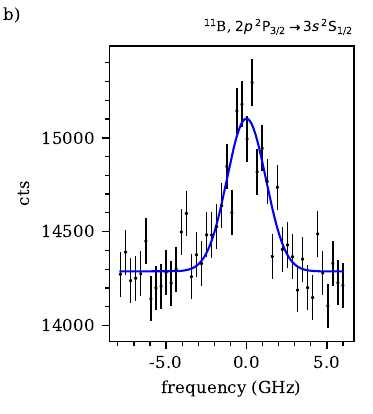}
    \label{fig:11BD2res}
    \end{subfigure}
    \begin{subfigure}[t]{0.33\textwidth}
    %\caption{$^{10}$B, from $^2\mathrm{P}_{\nicefrac{1}{2}}$}
    \centering
        \includegraphics{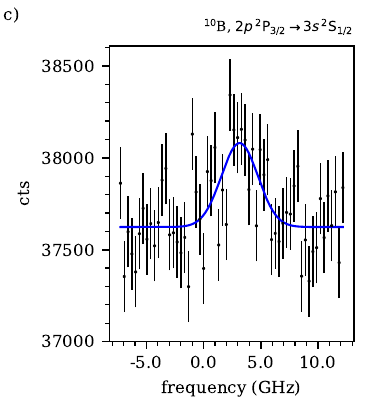}
        \label{fig:10Bres}
    \end{subfigure}
        \caption{Measured spectra in neutral boron. a) Resonance of the $2s^2\,2p\;^2\mathrm{P}_{\nicefrac{1}{2}} \rightarrow 2s^2\,3s\;^2\mathrm{S}_{\nicefrac{1}{2}}$ transition in $^{11}$B. The hyperfine structure of this transition is well known and has been additionally fitted by fixing the known hyperfine $A$ parameters of both levels and the expected intensity ratios of the hyperfine peaks leaving only the Gaussian width, the overall intensity and the center-of-gravity as free parameters. The small peaks are the individual hyperfine components and the yellow dashed line represents their sum. The latter is indistinguishable from the blue line, i.e., the simple Gaussian fit.  b) Resonance of the $2s^2\,2p\;^2\mathrm{P}_{\nicefrac{3}{2}} \rightarrow 2s^2\,3s\;^2\mathrm{S}_{\nicefrac{1}{2}}$ transition in $^{11}$B. 
        %The frequency axis is relative to \SI{1199900516}{\mega\hertz}. 
        c) Resonance of the $2s^2\,2p\;^2\mathrm{P}_{\nicefrac{3}{2}} \rightarrow 2s^2\,3s\;^2\mathrm{S}_{\nicefrac{1}{2}}$ transition in $^{10}$B. The frequency axis here is relativ to the isotope $^{11}$B. For details see text.
        %The frequency axis is relative to \SI{1199900516}{\mega\hertz}.
        }\label{fig:fits}
\end{figure*}

Despite these very low rates, we were able to record resonances of both $2s^2\,2p\,^2\mathrm{P}_{\nicefrac{1}{2},\nicefrac{3}{2}} \rightarrow 2s^2\,3s\,^2\mathrm{S}_{\nicefrac{1}{2}}$ fine-structure transitions in $^{11}$B (Fig.\,\ref{fig:fits}a,b) and a resonance of the $2s^2\,2p\ ^2\mathrm{P}_{\nicefrac{3}{2}} \rightarrow 2s^2\,3s\,^2\mathrm{S}_{\nicefrac{1}{2}}$ transition in $^{10}$B (Fig.\,\ref{fig:fits}c).
All transitions in $^{11}$B and $^{10}$B were recorded with $\SI{18}{\kilo \volt}$ acceleration voltage and additionally three measurements of the $2s^2\,2p\,^2\mathrm{P}_{\nicefrac{1}{2}} \rightarrow 2s^2\,3s\,^2\mathrm{S}_{\nicefrac{1}{2}}$ transition in $^{11}$B were taken with $\SI{13}{\kilo\volt}$ acceleration voltage. Depending on the acceleration voltage, the time gate was set to $0.5-\SI{3}{\micro\second}$ and $0.5-\SI{3.5}{\micro\second}$ after the detachment-laser pulse. \mycomment{\textcolor{red}{How does this correspond to the previous figure with 6 microseconds starting gate ?}}

It took about \SI{30}{\minute} to collect enough data for a single $^{11}$B spectrum and \SI{2}{\hour} to see a resonance in $^{10}$B, corresponding to their natural abundance ratio of 4:1. 
The spectra were fitted with a simple Gaussian profile since the hyperfine structure is unresolved and the peak does not exhibit any sign of substructure within its limited statistics. 
The Gaussian width parameter $\sigma$ is varying between $\SI{1000}{\mega\hertz}$ and $\SI{1500}{\mega\hertz}$. This agrees roughly with the hyperfine spectrum spreading across approximately $\SI{1300}{\mega\hertz}$ and $\SI{800}{\mega\hertz}$ for $^{11}$B and $^{10}$B, respectively \cite{Maa.2019} observed in high resolution spectra on a collimated thermal atomic beam. To estimate the systematic difference between the observed center of the Gaussian fit and the center of gravity of the underlying hyperfine structure, we performed a fit of the hyperfine structure in the D1 line of $^{11}$B, which is well known from previous measurements \cite{Maa.2019}. We fixed the hyperfine $A$ factors of the two levels to the known values \cite{Lew.1960,Maass.2020} and the peak intensity ratios according to the theoretical Racah intensities. The hyperfine splitting obtained in this way is illustrated in Fig.\,\ref{fig:fits}(a). The sum of the four Gaussians, indicated by the dashed line, has been found to be practically indistinguishable from the simple Gaussian fit. The center of gravity of the hyperfine structure is shifted with respect to the simple Gaussian center by about \SI{1.4}{\GHz}. The combination of a rather unstable acceleration potential, caused by feedback between the different power supplies required for the negative ion sputtering source, and the long integration times is probably the dominant contribution to the large resonance linewidth that completely smears out the hyperfine structure. 

We obtain a rough value for the isotope shift in the $2s^2\,2p\ ^2\mathrm{P}_{\nicefrac{3}{2}} \rightarrow 2s^2\,3s\;^2\mathrm{S}_{\nicefrac{1}{2}}$ transition from the difference of the measured transition frequencies. The weighted mean of seven recorded spectra was used for $^{11}$B, while only a single spectrum was recorded for $^{10}$B. The uncertainties are estimated as their standard deviation and the statistical fit uncertainty, respectively. Our uncertainty is dominated by the large scattering of the center positions in $^{11}$B, probably caused by the voltage fluctuations mentioned above, and the fitting uncertainty of $^{10}$B, being of similar size due to the bad statistics. This results in an isotope shift of \SI{-4.8(1.4)}{\giga\hertz}, which is in good agreement with the much more accurate measurement of \SI{-5.0313\pm 0.0020}{\GHz} reported in \cite{Maa.2019} for the  $2s^2\,2p\ ^2\mathrm{P}_{\nicefrac{1}{2}} \rightarrow 2s^2\,3s\;^2\mathrm{S}_{\nicefrac{1}{2}}$ fine-structure transition. This is expected since the splitting isotope shift, \textit{i.e.}, the difference in isotope shift between two fine-structure components, is typically of the order of a few MHz \cite{Drake.2005,Nortershauser.2015,Wang2017,Muller.2026}.

\section{Summary and conclusions}\label{sec:sum}
We demonstrated collinear laser spectroscopy on a beam of neutral B atoms, produced from photodetachment of B$^-$. A ToF measurement approach working with ion sources that deliver continuous ion beams was realized, which can be easily adapted to any collinear apparatus. Additionally, a Wien filter was implemented at the COALA setup to separate the boron ions from a large background of other ions leaving the sputter source.
The photodetachment method allows to neutralize the boron ion beam with a final population distributed only between the fine-structure components of the lowest atomic level, in our case the $2s^2 2p\ ^2\mathrm{P}_{1/2}$ and $2s^2 2p\ ^2\mathrm{P}_{3/2}$ fine-structure doublet. This eliminates the background signal from cascading decays as it would occur in the neutralization process happening in charge exchange cells.
With these preparations, we succeeded to take first spectra of a $^{10,11}$B neutral atomic beam generated by photodetachment of negative ions despite the very low yields caused dominantly by the very small duty cycle of the detachment laser and the small overlap between the ionization laser and the ion beam , which had a relatively large emittance.  
%Resonances of two transitions in $^{11}$B and one transition in $^{10}$B could be measured. 
While we have demonstrated the general technique, several improvements are required to improve efficiency and accuracy. Besides an improved voltage stabilization, particularly  higher neutralization efficiencies are crucial. Therefore, one could use a detachment laser with higher repetition rate and more power, two properties that are in conflict with each other. Better ion beam properties are required as well, \textit{i.e.}, a beam with less energy spread and better emittance. Both could be reached by combing the ion source with a radio-frequency quadrupole cooler and buncher \cite{Nieminen.2002}. In this case, the longitudinal as well as the transversal beam emittance will be drastically reduced, and ion bunches could be synchronized with the laser pulses. This will lead to a vast improvement in spatial as well as in temporal overlap with the detachment laser, and would even allow to use repetition rates of only a few 10 to 100\,Hz and have nevertheless 100\% temporal overlap. This is nowadays similarly applied in the collinear resonance ionization (CRIS) technique \cite{Flanagan.2013,Lynch.2014,Neugart.2017}. Switching additionally the cw laser beam with an electro-optical modulator, synchronous to the appearance of the short bunch of neutralized atoms at the optical detection region, will strongly reduce optical population transfer to non-resonant states upstream of the detection region. This has previously been demonstrated for the operation with a charge exchange cell \cite{Voss.2013}. 
With these improvements, the detachment process could be readily employed to obtain neutral fast beams with only ground-state population for efficient and accurate spectroscopy.

\section*{Acknowledgements}
This work was supported by the Deutsche Forschungsgemeinschaft (DFG, German Research Foundation) – Project-ID 279384907 – SFB 1245 and the German Federal Ministry for Education and Research (BMBF) under Contract No. 05P19RDFN1.

\appendix

\bibliographystyle{elsarticle-num} 
\bibliography{lit}

\end{document}